\documentclass[12pt]{iopart}

\usepackage{graphicx}
\usepackage{epstopdf}
\begin{document}

\title{Magneto-optical characterizations of FeTe$_{0.5}$Se$_{0.5}$ thin films with critical current density over 1 MA/cm$^2$ }

\author{Yue Sun$^{1*}$, Yuji Tsuchiya$^1$, Sunseng Pyon$^1$, Tsuyoshi Tamegai$^{1}$, Cheng Zhang$^{2}$, Toshinori Ozaki$^{2}$, Qiang Li$^{2}$}

\address{$^1$Department of Applied Physics, The University of Tokyo, 7-3-1 Hongo, Bunkyo-ku, Tokyo 113-8656, Japan

$^2$Condensed Matter Physics and Materials Science Department, Brookhaven National Laboratory, Upton, New York 11973, USA}

\begin{abstract}
We performed magneto-optical (MO) measurements on FeTe$_{0.5}$Se$_{0.5}$ thin films grown on LaAlO$_3$ (LAO) and Yttria-stabilized zirconia (YSZ) single-crystalline substrates. These thin films show superconducting transition temperature, \emph{T}$_c$ $\sim$ 19 K,  4 K higher than the bulk sample. Typical roof-top patterns can be observed in the MO images of thin films grown on LAO and YSZ, from which a large and homogeneous critical current density, \emph{J}$_c$, over 1 $\times$ 10$^6$ A/cm$^2$ at 5 K was obtained. Magnetic flux penetration measurement reveals that the current is almost isotropically distributed in the two thin films. Compared with bulk crystals, FeTe$_{0.5}$Se$_{0.5}$ thin film demonstrates not only higher \emph{T}$_c$, but also much larger \emph{J}$_c$, which is attractive for applications.
\end{abstract}

\maketitle

\section{Introduction}
After the discovery of superconductivity in iron-based superconductors (IBSs) in 2008 \cite{1}, extensive researches have revealed that this class of superconductors have fascinating fundamental properties for applications \cite{2}. The upper critical field, \emph{H}$_{c2}$, reaches over 50 T in all field directions, and the anisotropy is moderate, much smaller than the cuprate superconductors. Accompany to this small anisotropy, the superconducting transition under magnetic fields does not show an appreciable broadening, and the separation between \emph{H}$_{c2}$  and the irreversibility field is much narrower than cuprates \cite{3}. Among the family of IBSs, FeTe$_{1-x}$Se$_{x}$ has some practical advantages. Although the \emph{T}$_c$ is typically below 20 K, they exhibit lower anisotropy $\sim$ 2 with \emph{H}$_{c2}$ $\sim$ 50 T \cite{4,5}. On the other hand, its simple structure and less toxic nature are also preferable for fabricating wires and thin films, although the critical current density, \emph{J}$_c$, in FeTe$_{0.5}$Se$_{0.5}$ bulks and wires \cite{6,7} is still much lower than that of the Ba$_{0.6}$K$_{0.4}$Fe$_2$As$_2$ wires \cite{8}. Our previous reports have demonstrated that high-quality FeTe$_{0.5}$Se$_{0.5}$ thin films can maintain a large \emph{J}$_c$ over 10$^6$ A/cm$^2$ at 4.2 K. Even under the field of 30 T, \emph{J}$_c$'s are still exceeding 10$^5$ A/cm$^2$ \cite{9,10}.

For the practical application of superconducting wires and tapes, local characterizations of  \emph{J}$_c$ distribution are very important. Magneto-optical (MO) imaging has proven to be extremely powerful for such purposes. It can visualize the spatial distribution of \emph{J}$_c$, and is widely used to study the connectivity between grains in IBSs \cite{6, 11, 12, 13, 14, 15,16}. In this report, we carefully studied the critical current density and its distribution in FeTe$_{0.5}$Se$_{0.5}$ thin films by MO imaging for the first time. FeTe$_{0.5}$Se$_{0.5}$ thin films grown on LAO and YSZ shows a large, homogeneous, and almost isotropic \emph{J}$_c$ over 1 $\times$ 10$^6$ A/cm$^2$ at 5 K.

\section{Experimental details}
FeTe$_{0.5}$Se$_{0.5}$ thin films with thickness about 1000 {\AA} were grown by the pulsed laser deposition. Films were deposited on CeO$_2$ buffered single-crystalline substrates, LaAlO$_3$ (LAO) and Yttria-stabilized zirconia (YSZ). Details of the growth conditions have been reported in our previous publications \cite{17,18}. X-ray diffraction patterns reported in our previous paper show that only (00\emph{l}) peaks from the FeTe$_{0.5}$Se$_{0.5}$ thin films are present \cite{10}. The inset of Figure 1 shows a typical $\phi$ scan of the (101) peak from the film grown on YSZ, manifesting a sharp full width at half maximum (FWHM, $\Delta$$\phi$) of 1.3$^{\circ}$. The above results indicate the good in-plane alignments of FeTe$_{0.5}$Se$_{0.5}$ thin films without grain boundaries.

Resistivities were measured using the standard four-probe method. Magnetization measurements were performed using a commercial superconducting quantum interference device (SQUID). Magneto-optical (MO) images were obtained by using the local field-dependent Faraday effect. For MO measurements, the films were cut into rectangular shapes with dimensions about 1 $\times$ 1 mm$^2$ for that grown on LAO, and 0.85 $\times$ 0.78 mm$^2$ for that grown on YSZ.  An in-plane magnetized garnet indicator film was placed in direct contact with the surface of the sample and the image of the reflected light, which is related to the local magnetic induction around the sample, was captured by using an optical microscope and a cooled-CCD camera (ORCA-ER, Hamamatsu). The sample was cooled by a He-flow-type cryostat (Microstat-HR, Oxford Instruments). A sketch of the MO imaging system can be found in our previous report \cite{19}. To enhance the visibility and remove unnecessary contrasts coming from the defects of the garnet film, differential MO images were obtained by taking the difference between the two integrated images at \emph{H} = \emph{H}$_a$ and \emph{H} = 0 \cite{11,20}.

\section{Results and discussion}
\begin{figure}\center

　　\includegraphics[width=8cm]{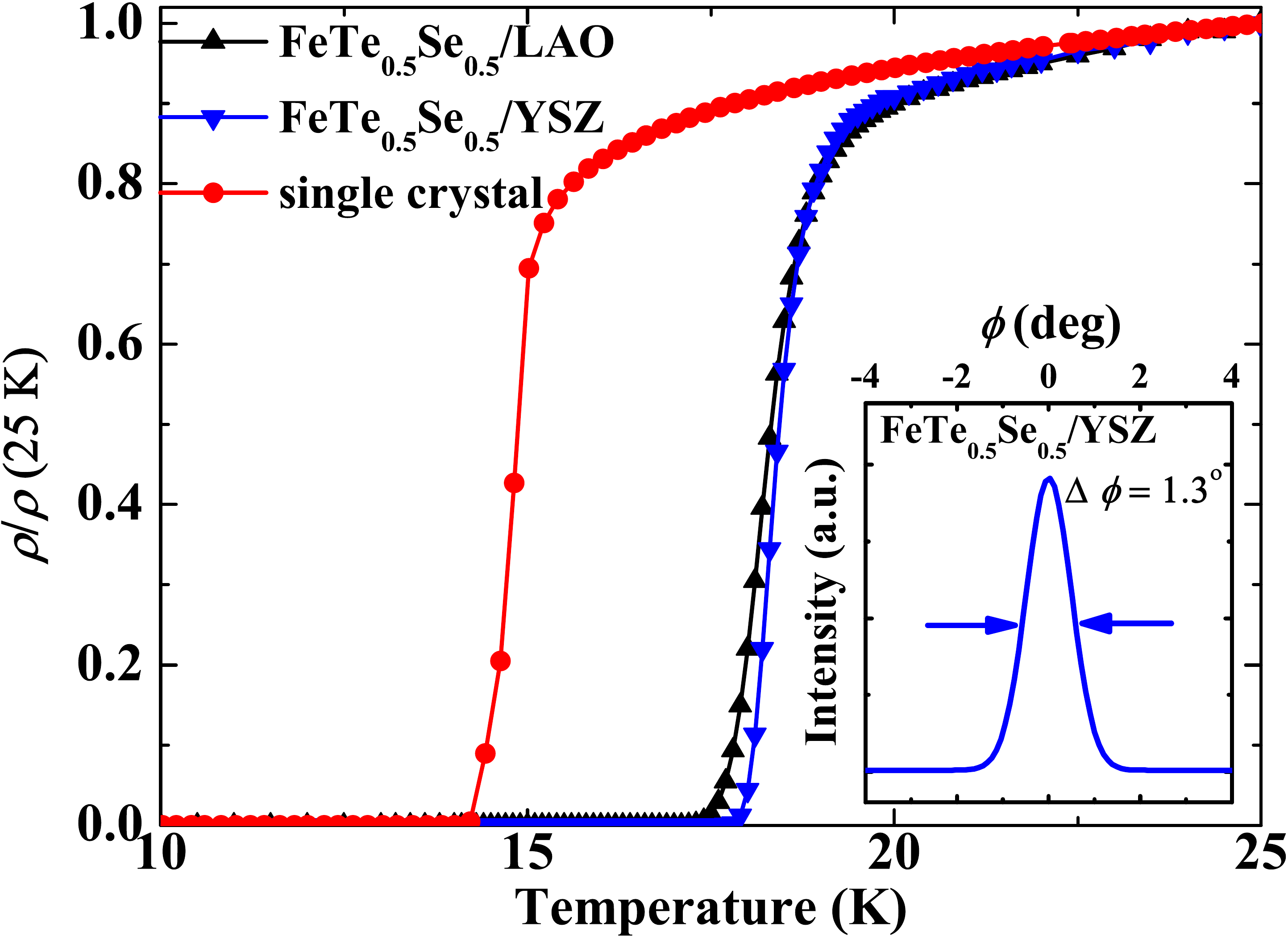}\\
　　\caption{ Resistive superconducting transition of  FeTe$_{0.5}$Se$_{0.5}$ thin films grown on LAO and YSZ substrates together with that of FeTe$_{0.6}$Se$_{0.4}$ single crystal. The inset shows the $\phi$ scan of (101) peak from the film grown on YSZ.}\label{}
\end{figure}

Figure 1 shows the resistive transitions of FeTe$_{0.5}$Se$_{0.5}$ thin films grown on LAO and YSZ substrates together with that of single crystal. \emph{T}$_c$ of the single crystal is $\sim$ 15 K, similar to the previous reports \cite{15,21}. The thin films grown on LAO and YSZ single-crystalline substrates both show \emph{T}$_c$ $\sim$ 19 K, 4 K higher than the bulk sample. The behavior of higher \emph{T}$_c$ observed in thin films than that of bulk samples is common in iron chalcogenides. Epitaxial strain in the films from substrates is thought to be a contributing factor for higher \emph{T}$_c$ observed in the superconducting iron chalcogenide films. \cite{17,18,22}.
\begin{figure}\center

　　\includegraphics[width=8cm]{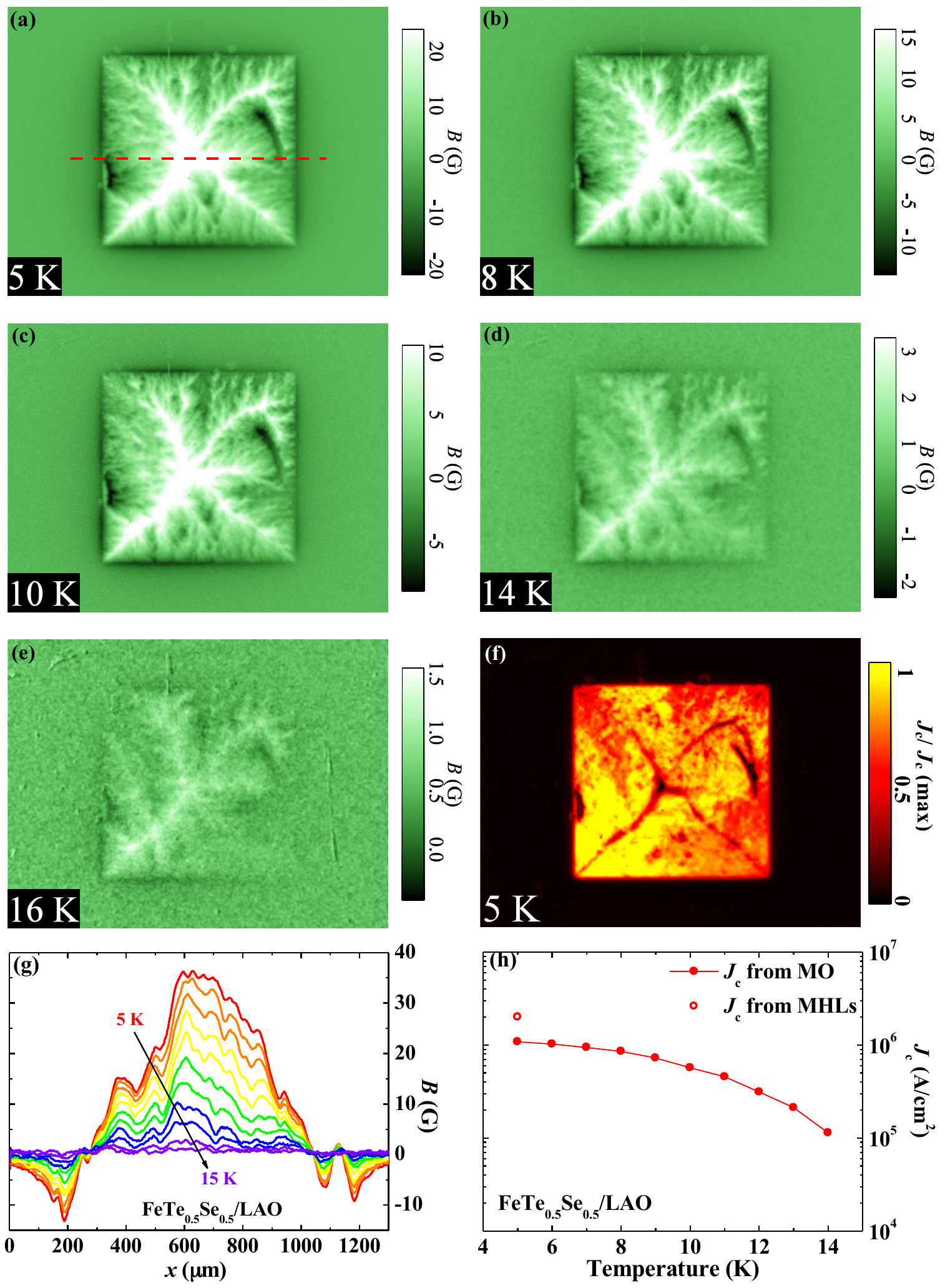}\\
　　\caption{MO images in the remanent state in FeTe$_{0.5}$Se$_{0.5}$ thin film grown on LAO substrate at (a) 5, (b) 8, (c) 10, (d) 14, (e) 16 K. (f) Spatial distribution of $|$\emph{J}$_c$$|$ at 5 K. (g) Local  magnetic induction profiles at different temperatures taken along the dashed lines in (a). (h)  Temperature dependence of \emph{J}$_c$ derived from MO, together with that obtained from MHLs (shown as the open circle).  }\label{}
\end{figure}

MO images in the remanent state are prepared by applying magnetic field, large enough to totally penetrate the sample, along \emph{c}-axis, and removing it after zero-field cooling. Typical MO images of FeTe$_{0.5}$Se$_{0.5}$ thin film grown on LAO substrate from 5 to 16 K are shown in Figure 2(a) - (e). The MO image manifests a typical roof-top pattern, similar to that observed in high-quality FeTe$_{0.6}$Se$_{0.4}$ single crystal \cite{15, 23,24,25,26}, indicating a nearly uniform current flow in the thin film. To directly observe the distribution of \emph{J}$_c$ in the film, we convert the MO images into the current distribution with a thin-sheet approximation by using the fast Fourier transform process to simplify the Biot-Svart's law \cite{27,28}. A typical image of the distribution of the modulus of \emph{J}$_c$ at 5 K is shown in Figure 2(f), which again manifests that the current is homogeneously distributed in most parts of the thin film. Figure 2(g) shows the profiles of the magnetic induction along the dashed line in Figure 2(a) at different temperatures. From this profile, the critical current density, \emph{J}$_c$, for the thin film can be roughly estimated from the following formula
\begin{equation}
\label{eq.1}
 \Delta B = \frac{8J_c\cdot2d}{c}\int^{a}_0 \frac{\omega^2}{\sqrt{z_g^2+2\omega^2}(z_g^2+\omega^2)}d\omega,
\end{equation}
where $\Delta$\emph{B} is the trapped field in the film, $2d$ and $2a$ are the thickness and length/width of the sample, respectively, and $z_g$ is the distance from the film surface to MO indicator, and $c$ is the velocity of light, which is equal to 10 in the practical unit (derivation of the formula can be seen in the supplement). Substituting $z_g$ = 3 $\mu$m from our experience in MO imaging \cite{29}, and $a$ = 500 $\mu$m into the above formula, the integral part can be numerically calculated as 3.471. Thus, magnetic field along $z$-axis can be simply estimated as: $J_c$ = $\Delta B$/($2d\times$2.78).
Temperature dependence of \emph{J}$_c$ obtained from MO image is plotted in Figure 2(h). \emph{J}$_c$ at 5 K shows a large value of $\sim$ 1.1 $\times$ 10$^6$ A/cm$^2$, which is close to that obtained from magnetic hysteresis loops (MHLs) by the extended Bean model \cite{30} (shown as the open circle in Figure 2(h))
\begin{equation}
\label{eq.2}
J_c=20\frac{\Delta M}{2a(1-a/3b)},
\end{equation}
where $\Delta$\emph{M} is \emph{M}$_{down}$ - \emph{M}$_{up}$, \emph{M}$_{up}$ [emu/cm$^3$] and \emph{M}$_{down}$ [emu/cm$^3$] are the magnetization when sweeping fields up and down, respectively, 2$a$ [cm] and 2$b$ [cm] are sample widths (\emph{a} $<$ \emph{b}).
\begin{figure}\center

　　\includegraphics[width=8cm]{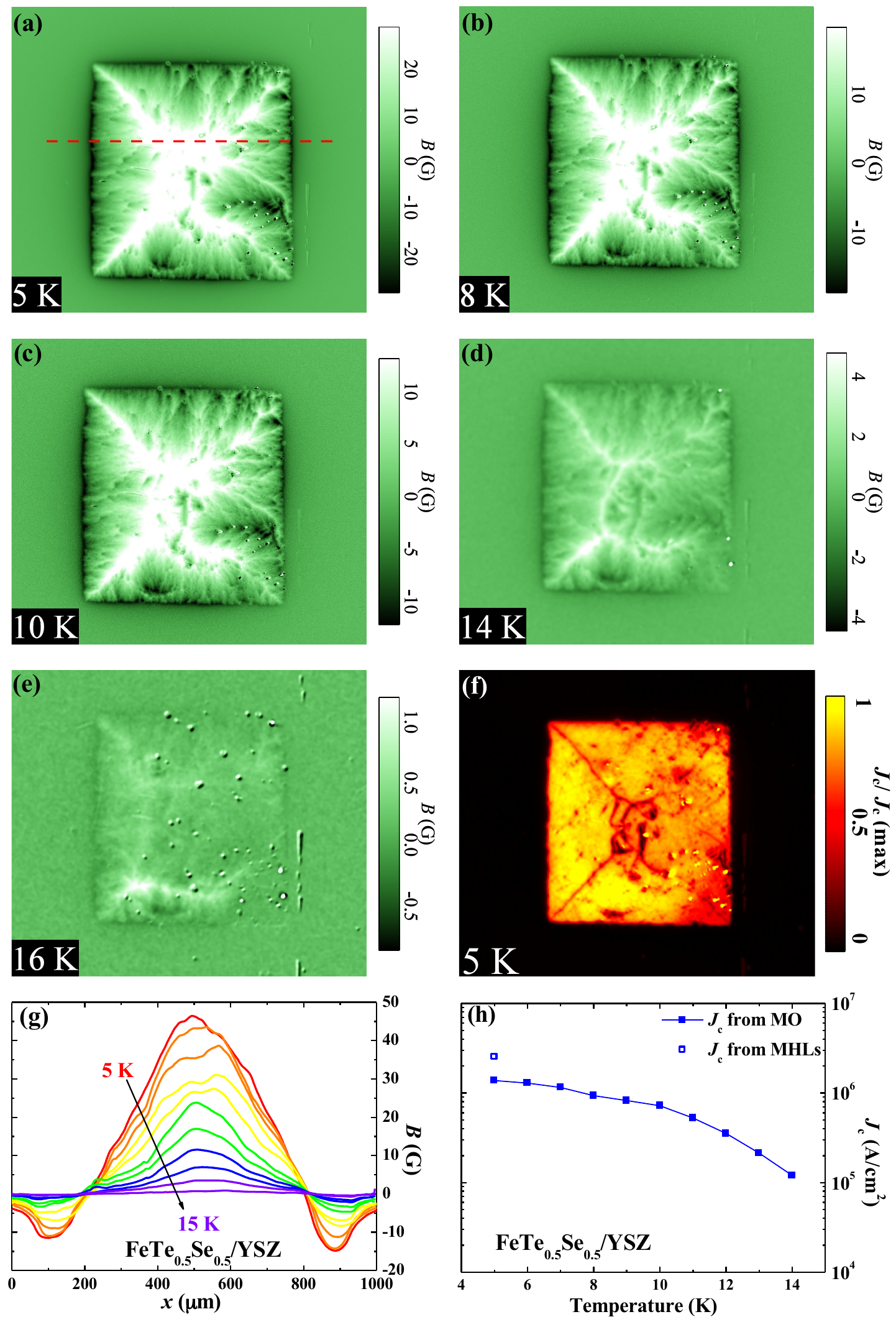}\\
　　\caption{MO images in the remanent state in FeTe$_{0.5}$Se$_{0.5}$ thin film grown on YSZ substrate at (a) 5, (b) 8, (c) 10, (d) 14, (e) 16 K. (f) Spatial distribution of $|$\emph{J}$_c$$|$ at 5 K. (g) Local  magnetic induction profiles at different temperatures taken along the dashed lines in (a). (h)  Temperature dependence of \emph{J}$_c$ derived from MO, together with that obtained from MHLs (shown as the open square). }\label{}
\end{figure}

Figure 3(a) - (e) show the MO images in the remanent state of FeTe$_{0.5}$Se$_{0.5}$ thin film grown on YSZ substrate at temperature ranging from 5 to 16 K. Similar to the film grown on LAO substrate, MO images at temperatures lower than 14 K show typical roof-top patterns, indicating the homogeneous distribution of critical current density in the thin film grown on YSZ substrate. The homogeneous distribution of \emph{J}$_c$ can be directly observed in its spatial distribution as shown in Figure 3(f). The superconductivity observed above 15 K seems less homogeneous because just parts of the thin film can trap field, as shown in Figure 3(e). Profiles of the magnetic induction along the dashed line in Figure 3(a) at different temperatures are shown in Figure 3(g). \emph{J}$_c$'s at different temperatures are roughly estimated and shown in Figure 3(h). The value of \emph{J}$_c$ reaches $\sim$ 1.4 $\times$ 10$^6$ A/cm$^2$ at 5 K, and is also close to the global \emph{J}$_c$ obtained from MHLs shown as the open square in Figure 3(h).
\begin{figure}\center
　　\includegraphics[width=15cm]{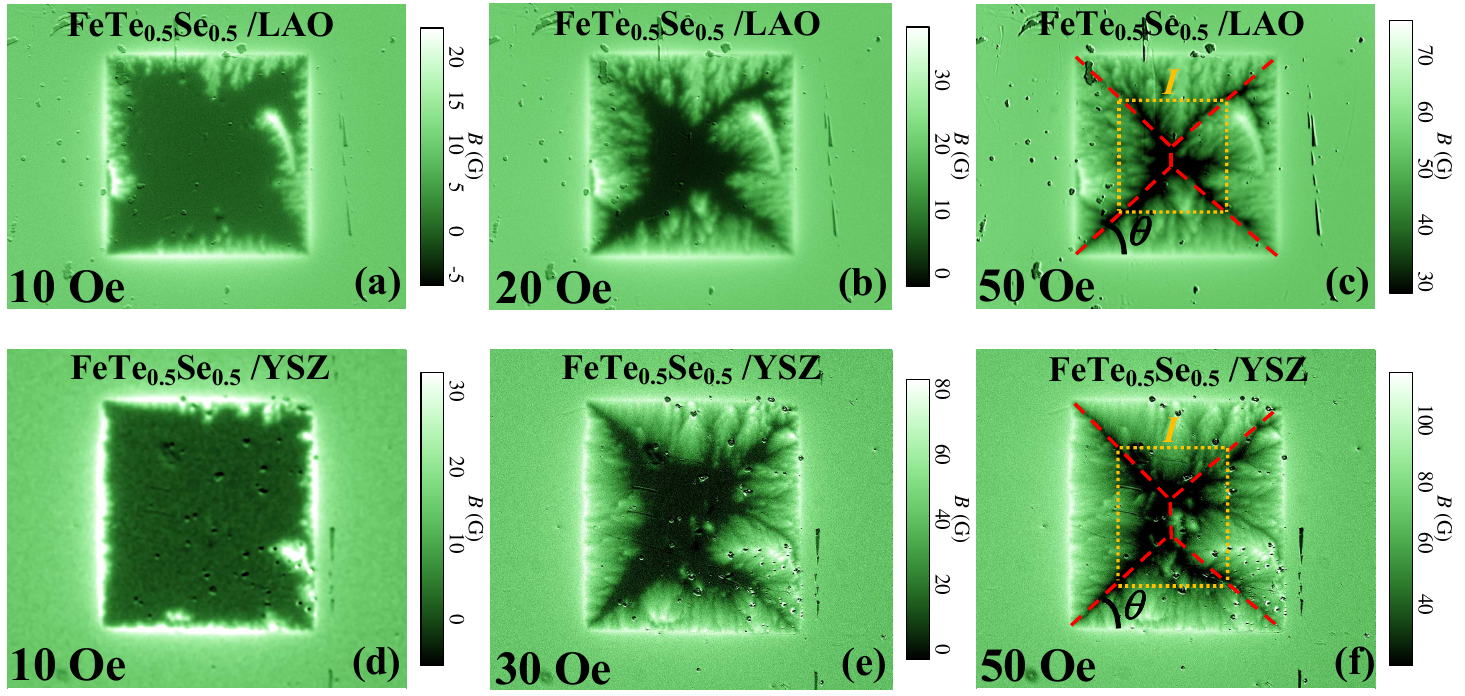}\\
　　\caption{Magneto-optical images of flux penetrations into the FeTe$_{0.5}$Se$_{0.5}$ thin films grown on LAO and YSZ substrates at 5 K under different magnetic fields. The red dashed lines in (c) and (f) show the current discontinuity lines, which cannot be crossed by vortices. The yellow dotted lines are the schematics of the current distributions in the thin films. }\label{}
\end{figure}

Figure 4(a) - (c) and (d) - (f) reveal the penetration of magnetic flux at 5 K for FeTe$_{0.5}$Se$_{0.5}$ thin films grown on LAO and YSZ, respectively. Obviously, flux gradually penetrate the two thin films with increasing applied field. More importantly, the typical current discontinuity lines (so-called \emph{d}-line), which cannot be crossed by vortices, can be directly observed and marked by the dashed line in Figure 4(c) and (f). By measuring the angles of the discontinuity line for the rectangular film, the in-plane anisotropy of the current densities can be easily estimated \cite{31}. In this case, the angle $\theta$ is close to 45$^{\circ}$, indicating that the critical current density in the \emph{ab}-plan is almost isotropic. Schematics of the current distributions is presented by the dotted lines.
\begin{figure}\center

　　\includegraphics[width=8cm]{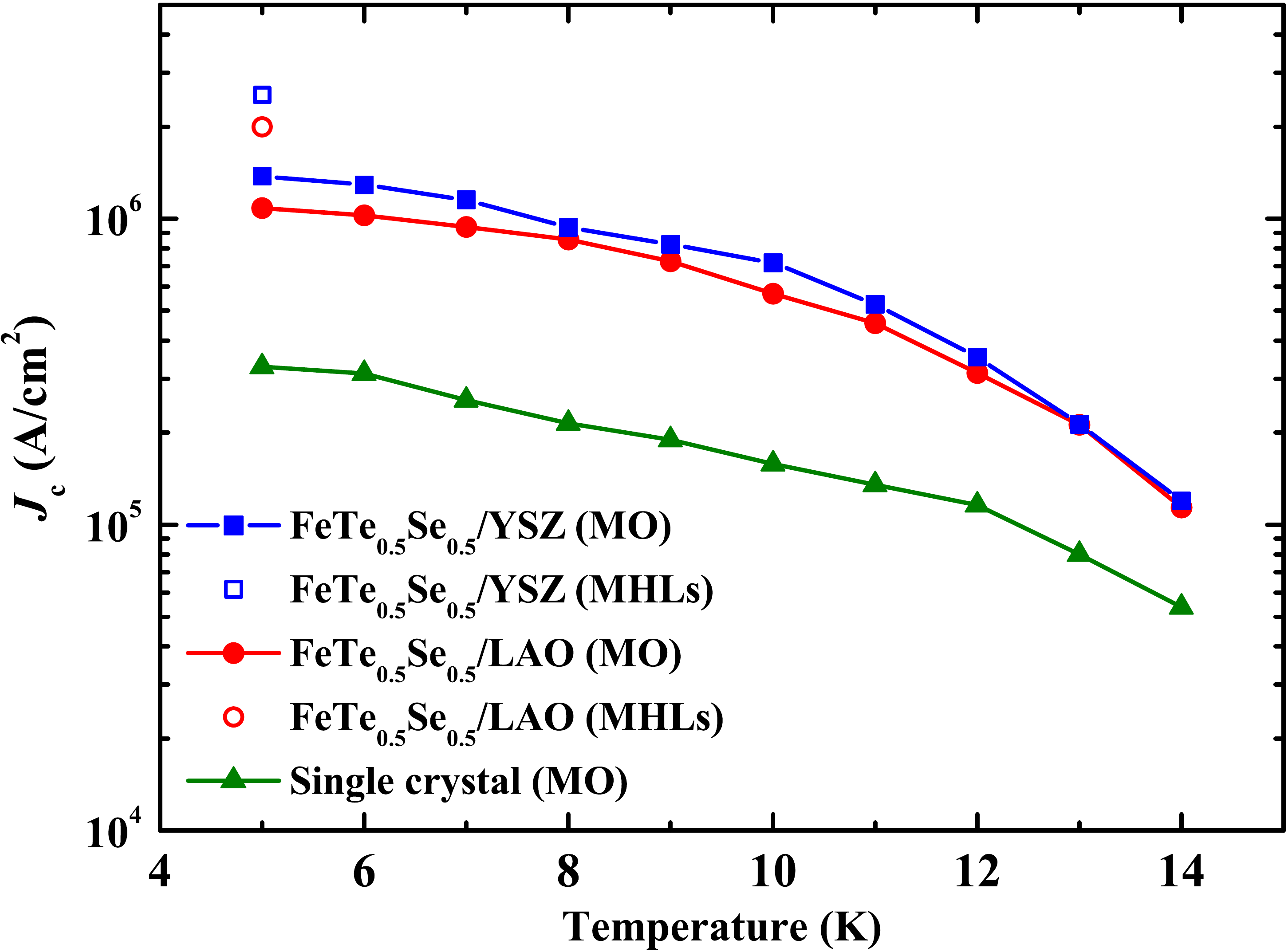}\\
　　\caption{Temperature dependence of \emph{J}$_c$ obtained from MO and MHLs for FeTe$_{0.5}$Se$_{0.5}$ thin films grown on LAO and YSZ substrates, together with that of FeTe$_{0.6}$Se$_{0.4}$ single crystal[15].}\label{}
\end{figure}

Here, we should point out that we have performed extensive structural and chemical analysis of these films at room temperature by both synchrotron-based x-ray diffraction and analytical transmission electron microscope (TEM) equipped with nano-probe Electron Energy Loss Spectroscopy (EELS). We found no evidence of structural and chemical inhomogeneities of these films at macroscopic level. However, we did notice some minor inhomogeneities in the trapped magnetic flux patterns from MO images. Although the observed inhomogeneities are too small to have meaningful impact to our determination of the local critical current density of these films, they do suggest that these films are very sensitive to small perturbations, which are presumably related to the strain built in the films. There are several possible ways to induce strain in thin films. Among them, are epitaxial strain due to slightly lattice constant mismatch between the film and substrates, strain induced during the sample cool-down from processing temperature (400$^{\circ}$C) to room temperature or from room temperature to low temperature for MO experiments. The latter type of thermally induced strain tends to be less macroscopically uniform, and hence may cause variation of \emph{T}$_c$ and \emph{J}$_c$ across the entire film, that might affect the uniformity of the flux patterns

In Figure 5, we compare the temperature dependence of \emph{J}$_c$'s for FeTe$_{0.5}$Se$_{0.5}$ thin films grown on LAO and YSZ substrates together with that from high-quality single crystal, which also manifests a roof-top pattern in MO images \cite{15,23}. The values of \emph{J}$_c$'s for thin films are over 10$^6$ A/cm$^2$, one order of magnitude higher than that of the single crystal. The above results show that FeTe$_{0.5}$Se$_{0.5}$ thin films exhibit not only higher \emph{T}$_c$, but also larger values of \emph{J}$_c$ compared with the bulk sample, which is very attractive for applications. This values of \emph{J}$_c$ in FeTe$_{0.5}$Se$_{0.5}$ thin films are close to those of the Nb$_3$Sn wires. Considering the much lower \emph{H}$_{c2}$ of Nb$_3$Sn, FeTe$_{0.5}$Se$_{0.5}$ may be a respectable candidate for applications at the liquid helium temperatures. In addition, we already reported the successful fabrication of coated FeTe$_{0.5}$Se$_{0.5}$ tapes with \emph{J}$_c$ over 10$^6$ A/cm$^2$ under zero field, and over 1 $\times$ 10$^5$ A/cm$^2$ under 30 T magnetic field at 4.2 K \cite{10}. All these results show that the FeTe$_{0.5}$Se$_{0.5}$ is suitable for high-field applications.

\section{Conclusions}
In conclusion, MO images were taken on FeTe$_{0.5}$Se$_{0.5}$ thin films grown on LAO and YSZ single-crystalline substrates. MO images at remanent state show typical roof-top patterns. Large, homogeneous, and almost isotropic \emph{J}$_c$ over 1 $\times$ 10$^6$ A/cm$^2$ at 5 K were obtained. The properties of large and field-independent \emph{J}$_c$, very high upper critical field, and relatively low cost make FeTe$_{0.5}$Se$_{0.5}$ an attractive candidate for applications at liquid helium temperatures, especially under high magnetic field.

\ack
Y.S. gratefully appreciates the support from Japan Society for the Promotion of Science. The work at the University of Tokyo was supported by Japan-China Bilateral Joint Research Project by the Japan Society for the Promotion of Science. The work at Brookhaven National Laboratory was supported by the US Department of Energy, Office of Basic Energy Science, Materials Sciences and Engineering Division, under contract no. DEAC0298CH10886.
\\
\\$^{*}$sunyue.seu@gmail.com

\newpage

\title{Supplementary Information: Estimation of  Critical Current Density from Magneto-optical Images}

\author{Yue Sun$^{1*}$, Yuji Tsuchiya$^1$, Sunseng Pyon$^1$, Tsuyoshi Tamegai$^{1}$, Cheng Zhang$^{2}$, Toshinori Ozaki$^{2}$, Qiang Li$^{2}$}

\address{$^1$Department of Applied Physics, The University of Tokyo, 7-3-1 Hongo, Bunkyo-ku, Tokyo 113-8656, Japan

$^2$Condensed Matter Physics and Materials Science Department, Brookhaven National Laboratory, Upton, New York 11973, USA}

\maketitle
To calculate critical current density, $J_c$, from the magneto-optical (MO) imaging, we consider the situation where homogeneous current flows in a square sample with dimensions $2a\times$$2a\times$2$d$, where $2a$ and $2d$ are the length (width) and thickness of the sample, respectively (figure 6). For our FeTe$_{0.5}$Se$_{0.5}$ thin films, $2a$ = 1000 $\mu$m, and $2d$ = 0.1 $\mu$m. Then we calculate the $z$-axis component of magnetic induction, $B_z$(0, 0, $z_g$), at the center of the sample and $z = z_g$ above the surface. The current is homogeneously distributed, and its direction is shown by thick blue arrows in figure 6. In this case, the sample can be divided into eight parts. Because of the symmetry, all the eight parts contribute equally to $B_z$(0, 0, $z_g$).
\begin{figure}\center

　　\includegraphics[width=8cm]{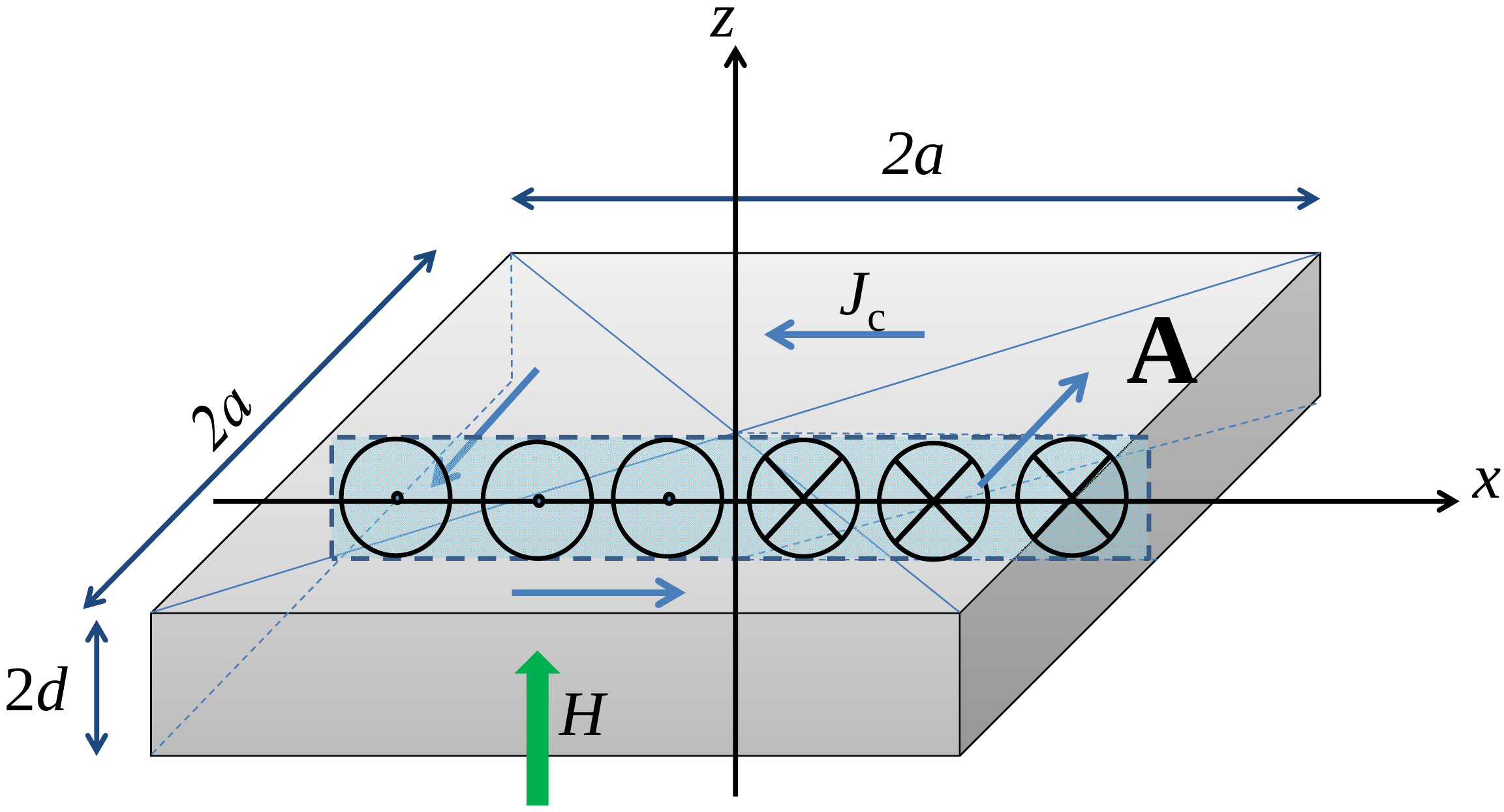}\\
　　\caption{Sketch of the square sample for calculation.}\label{}
\end{figure}

We first consider the region A, which is one of eight regions of the sample. In this region, a thin strip of current flows at a position $x$ = $\omega$ $\sim$ $\omega + d\omega$, as shown in figure 7. We assume that the sample is infinitely thin. The contribution of this region to $B_z$(0, 0, $z_g$) can be calculated by summing up all contributions from thin strip currents at $x = \omega$ to $\omega$ + $d\omega$ (figure 7) from $\omega = 0$ to $\omega = a$. Based on Biot-Savart law:
\begin{figure}\center

　　\includegraphics[width=8cm]{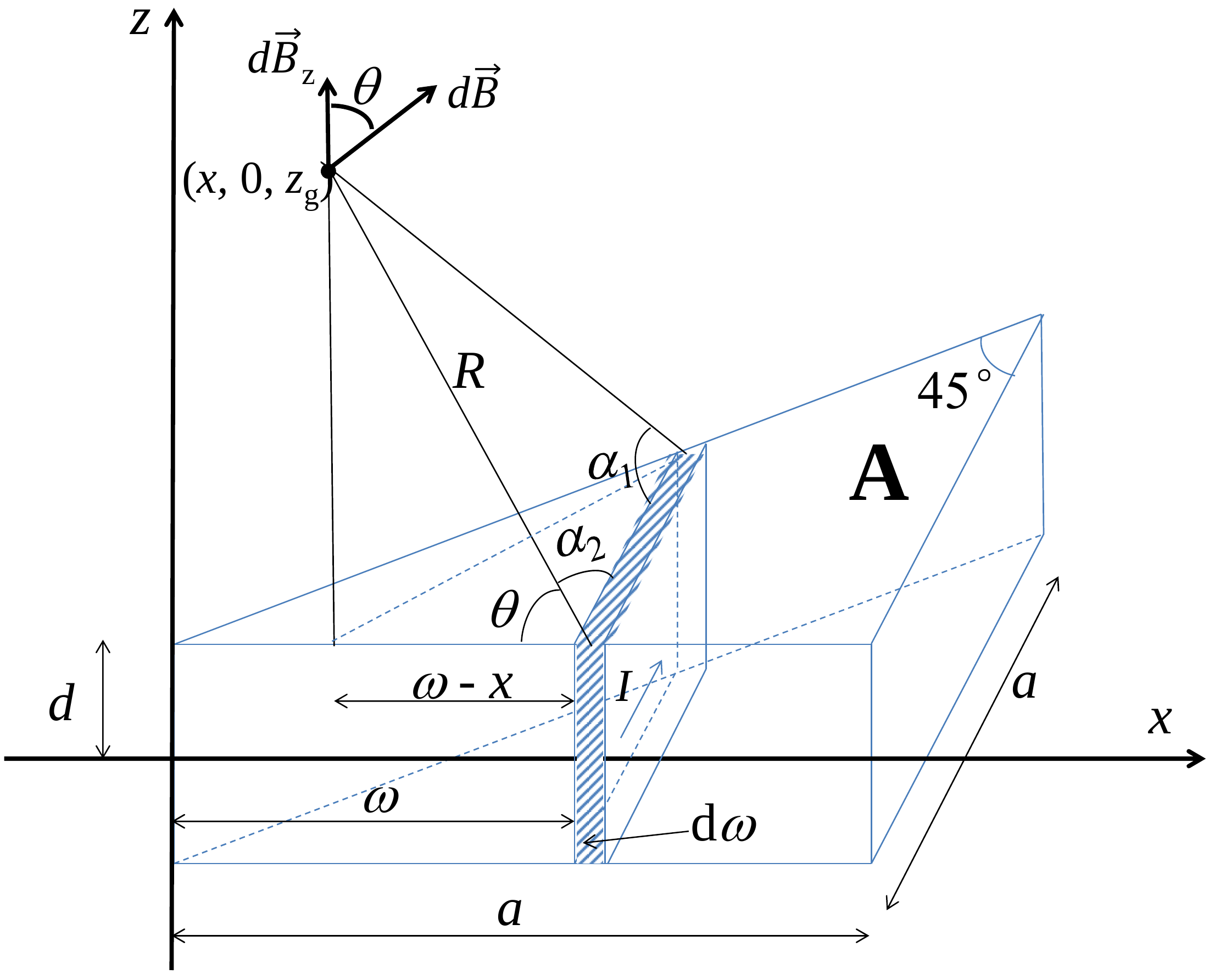}\\
　　\caption{Sketch of the enlarged region A.}\label{}
\end{figure}
\begin{equation}
\label{eq.1}
d\vec{B} = \frac{\mu_0}{4\pi}\int\frac{Id\vec{l} \times \vec{r}}{|\vec{r}|^3},
\end{equation}
where $d\vec{B}$ is the magnetic induction, $I d\vec{l}$ is the infinitely small current element of the thin strip carrying current $I$, $d\vec{r}$ is the vector from $d\vec{l}$ to the field point, the magnetic induction created by this strip of current can be expressed as
\begin{equation}
\label{eq.2}
dB = |d\vec{B}| = \frac{\mu_0I}{4\pi R}\int^{\alpha_2}_{\alpha_1} \sin\alpha d\alpha=\frac{\mu_0I}{4\pi R}(\cos\alpha_1-\cos\alpha_2).
\end{equation}
Since $\alpha_2$ = 90$^\circ$ and $\cos$ $\alpha_2$ = 0,
\begin{equation}
\label{eq.3}
dB=\frac{\mu_0I}{4\pi R}\cos\alpha_1.
\end{equation}
Thus, $z$-axis component of magnetic induction generated by this thin current strip can be written as
\begin{equation}
\label{eq.4}
dB_z =\frac{\mu_0J_c\cdot2d}{4\pi R}\cos\alpha_1\cos\theta d\omega,
\end{equation}
where $\theta$ is the angle between $d\vec{B}$ and $z$-axis, which can be expressed as
\begin{equation}
\label{eq.5}
\cos\theta = \frac{\omega-x}{R},
\end{equation}
where $R$ = $\sqrt{z_g^2+(\omega-x)^2}$ is the distance between the thin strip and the measured point. From the sketch in figure 7, we can easily obtain the following expression:
\begin{equation}
\label{eq.6}
\cos\alpha_1 = \frac{\omega}{\sqrt{z_g^2+2\omega^2+x^2-2\omega x}}.
\end{equation}
Combining equations (4) - (6), the magnetic field contributed from region A can be obtained as the integration from $\omega$ = 0 to $a$:
\begin{equation}
 B^A_z(x, 0, z_g) = \frac{\mu_0J_c\cdot2d}{4\pi}\int^a_0 \frac{\cos\alpha_1 \cos\theta}{R}dw
\end{equation}
\begin{equation}
\hspace{22mm}= \frac{\mu_0J_c\cdot2d}{4\pi}\int^a_0 \frac{\omega}{\sqrt{z_g^2+2\omega^2+x^2-2\omega x}}\frac{(\omega-x)}{[z_g^2+(\omega-x)^2]}d\omega.
\end{equation}
Then at the center of the sample, $x$ = 0,
\begin{equation}
 B^A_z(0, 0, z_g) = \frac{\mu_0J_c\cdot2d}{4\pi}\int^a_0 \frac{\omega^2}{\sqrt{z_g^2+2\omega^2}(z_g^2+\omega^2)}d\omega.
\end{equation}
 Since all the other 7 parts contribute equally as region A, the total magnetic induction along $z$-axis at $x$ = 0 can be expressed as:
\begin{equation}
 B_z(0, 0, z_g) = \frac{2\mu_0J_c\cdot2d}{\pi}\int^a_0 \frac{\omega^2}{\sqrt{z_g^2+2\omega^2}(z_g^2+\omega^2)}d\omega.
\end{equation}
 The above formula is derived in MKS unit, which can be simply reformed into cgs unit by replacing $\mu_0$ as 4$\pi/c$ ($c$ is the velocity of light, which is equal to 10 in the practical unit)
\begin{equation}
 B_z(0, 0, z_g) = \frac{8J_c\cdot2d}{c}\int^a_0 \frac{\omega^2}{\sqrt{z_g^2+2\omega^2}(z_g^2+\omega^2)}d\omega.
\end{equation}
Substituting $z_g$ = 3 $\mu$m from our experience in MO imaging \cite{1} (In this ref., we pushed the indicator hardly to the sample, thus the distance is reduced to $\sim$ 1.5 $\mu$m. Usually, the distance $z_g$ is about 2 $\sim$ 3 $\mu$m) and $a$ = 500 $\mu$m into the above formula, the integral part can be numerically calculated as 3.471.
Thus, magnetic induction along $z$-axis can be simply estimated as:
\begin{equation}
 B_z(0, 0, z_g) = \frac{8J_c\cdot2d}{c}\times3.471 = 2.78 \times J_c\cdot2d.
\end{equation}
Based on the calculation above, $J_c$ in our square sample can be estimated by using $\Delta B$ $\equiv$ $B_z(0, 0, z_g)$ as:
\begin{equation}
 J_c = \frac{1}{2.78}\frac{\Delta B}{2d}.
\end{equation}

Actually, in a two-dimensional superconducting strip in $xy$-plane with a width $2a$, magnetic induction along $z$-axis, $B_z(x)$ can be simply obtained by the same method, and has already been applied to estimate $J_c$ from MO measurements, as reported by Prozorov \cite{2} and our previous publication \cite{3}:
 \begin{equation}
 B_z(x) = \frac{J_c\cdot2d}{c}\ln{\frac{[z_g^2+(x-a)^2][z_g^2+(x+a)^2]}{(z_g^2+x^2)^2}}.
\end{equation}
 From this formula, $J_c$ can be estimated by
\begin{equation}
 J_c = \frac{1}{2.05}\frac{\Delta B}{2d}.
\end{equation}
The above result manifests that the prefactor only changes from 1/2.78 to 1/2.05 when the shape of the sample changes from square to thin strip. Since our FeTe$_{0.5}$Se$_{0.5}$ thin films are close to the case of square, equation (13) is adopted to estimate $J_c$.

\end{document}